\documentclass[manuscript, anonymous=false,authordraft,review=false]{acmart}

\usepackage{fontawesome}
\usepackage{float}
\usepackage{tabularx}
\usepackage{dcolumn} 
\newcolumntype{d}[1]{D{.}{.}{#1}}
\usepackage{float}
\usepackage{subcaption}
\usepackage{makecell}
 \usepackage{threeparttable} 
\usepackage{todonotes}
\usepackage[normalem]{ulem}
\usepackage{tikz}
\usepackage{adjustbox}
\usepackage{footmisc}

\newcommand{\blackcircled}[1]{%
  \tikz[baseline=(char.base)]{
    \node[circle,fill=black,text=white,inner sep=.5pt] (char) {\textbf{\sffamily #1}};
  }%
}

\useunder{\uline}{\ul}{}
\let\xtodo\todo
\renewcommand{\todo}[1]{\xtodo[inline,color=green!50]{#1}}

\definecolor{colortodo}{rgb}{0.9, 0.6, 0}    
\definecolor{colordaniel}{rgb}{0, 0.7, 0.35}
\definecolor{colorthomas}{rgb}{0.2, 0.4, 0.8}
\definecolor{colorrobin}{rgb}{0.2, 0.8, 0.9}
\definecolor{colordaniela}{rgb}{0.8, 0.2, 0.7}
\definecolor{colornotfinal}{rgb}{0.95, 0.2, 0.2} 
\definecolor{colorlev}{rgb}{0.4, 0.2, 0.2}

\AtBeginDocument{%
  }

\acmDOI{}
\acmISBN{}
\acmYear{}
\setcopyright{none}
\acmConference[]{}{}{}
\begin{document}

\title{Explaining Too Much? Understanding How Large Language Model Reasoning Traces Influence Performance and Metacognition}


\author{Daniela Fernandes}
\email{daniela.dasilvafernandes@aalto.fi}
\orcid{0009-0006-1332-7485}
\affiliation{%
  \institution{Aalto University}
  \city{Espoo}
  \country{Finland}
}

\author{Daniel Buschek}
\email{daniel.buschek@uni-bayreuth.de}
\orcid{0000-0002-0013-715X}
\affiliation{%
  \institution{University of Bayreuth}
  \city{Bayreuth}
  \country{Germany}
}

\author{Lev Tankelevitch}
\email{levt@microsoft.com}
\orcid{0000-0003-1286-5194}
\affiliation{%
  \institution{Microsoft Research Cambridge}
  \city{Cambridge}
  \country{United Kingdom}
}

\author{Thomas Kosch}
\email{thomas.kosch@hu-berlin.de}
\orcid{0000-0001-6300-9035}
\affiliation{%
  \institution{HU Berlin}
  \city{Berlin}
  \country{Germany}
}

\author{Robin Welsch}
\email{robin.welsch@aalto.fi}
\orcid{0000-0002-7255-7890}
\affiliation{%
  \institution{Aalto University}
  \city{Espoo}
  \country{Finland}
}

\renewcommand{\shortauthors}{Fernandes et al.}

\begin{abstract}

Large Language Model interfaces are increasingly verbose, exposing intermediate reasoning traces alongside final answers. Traces are framed as transparency mechanisms, yet it is unclear how people use them to solve problems. We report a preregistered between-subjects study (N = 559) in which participants solved ten LSAT-style reasoning problems under one of three conditions: an \textit{Answer-only} baseline, a \textit{Full-trace} revealed before the answer, and a \textit{Summary-trace} presented alongside the answer. Summaries preserved task performance at the no-trace baseline while significantly elevating trust and hedonic appeal, establishing that trace exposure shifts subjective appraisal of the interaction without bringing performance benefits. Under an open-weight reasoning model exposing verbose intermediate output, full traces additionally impaired performance relative to the answer-only baseline. Across all conditions, participants substantially overestimated their performance, and no trace format supported calibrated self-evaluation. Further analysis indicates that hedonic appeal, not trust, carries the indirect path to overestimation, consistent with a processing-fluency account. Reasoning traces are best understood as user-facing interface artifacts rather than transparent windows into model cognition, and calibration is unlikely to emerge from the traces themselves and may best be scaffolded by interactions that elicit users' own reasoning first.

\end{abstract}



\received{}
\received[revised]{ }
\received[accepted]{ }

\maketitle

\section{Introduction}
Large Language Models (LLMs) are becoming increasingly verbose, with users now exposed not only to final outputs but also to intermediate ``reasoning traces'', chain-of-thought-style outputs that resemble the model's inner workings \cite{barez2025chain}.
Showing reasoning traces carries the promise of transparency. They are meant to help users make sense of AI answers and scaffold complex workflows \cite{6645235,anthropic2025extended,openai2024learning,openai2025monitorability}. Yet prior research in XAI and HCI reports mixed findings. High verbosity can induce an illusion of understanding (e.g., illusions of knowledge \cite{fisher2021harder}), and fluent justifications may lead users toward overreliance even when the underlying conclusion is wrong \cite{10.1145/3397481.3450639, 10.1145/3449287}. Moreover, a model's ``chain-of-thought'' is not always faithful to its intermediate steps and can function as a polished, but misleading, justification \cite{10.5555/3666122.3669397,lindsey2025biology, barez2025chain}.

Despite this, providers explicitly market reasoning traces as tools for transparency \cite{lindsey2025biology}, and while generating these traces is costly, traces are billed at output rates and largely increase the amounts of tokens as compared to a direct answer \cite{sui2025stop, anthropic2025extended}. The design question is thus whether exposing reasoning traces delivers the transparency providers claim, with benefits for its users.

At the same time, while AI has the potential to improve task performance \cite{10.1145/3411764.3445717, steyvers_bayesian_2022,zulfikar_memoro_2024}, users often believe they perform better when assisted by AI \cite{FERNANDES2026108779, kloft2023ai, villa2023placebo} and struggle to align confidence with correctness. This highlights the core challenge of understanding not only how people can make better decisions with AI, but how AI interfaces shape \textit{metacognition}, that is, users' ability to monitor and evaluate their own decisions and regulate reliance during interaction with AI.%

In this paper, we investigate how the shape of reasoning traces affects performance and metacognition in human-AI reasoning contexts. 
More specifically, we ask (1) what the benefits and drawbacks of different reasoning-trace formats are, (2) how these formats affect task performance, (3) how they shape metacognition, and (4) whether they can support calibrated reliance on AI. 

To address these questions, we present findings from a preregistered between-subjects experiment (N = 559) in which participants solved ten LSAT-style logical reasoning problems with AI assistance under one of three interface conditions that span the practically relevant design space for exposing model reasoning:

\begin{itemize}
    \item \textit{Answer-only} condition (N = 187): An answer-only baseline that reflects early LLM interfaces.
    \item \textit{Full-trace} condition (N = 183): A full reasoning trace shown before the model's final answer, which remained hidden behind a button press. This ensured that participants engaged with the trace itself rather than the verdict, and mirrored the verbose outputs of current open-weight reasoning models.
    \item \textit{Summary-trace} condition (N = 189): A summary of reasoning traces presented together with the answer, of the kind now adopted by several commercial systems (see \cite{openai2024learning, anthropic2025extended}).
\end{itemize}

These three formats capture the dominant ways in which reasoning is currently surfaced to users and allow us to control verbosity in our study. By isolating the \textit{content} of the trace from how it is revealed, we can ask whether traces help, hurt, or need to be calibrated, and at what level of verbosity, if any, they support metacognitive accuracy rather than merely inflating the appearance of understanding. LSAT-style reasoning has been established as a robust measure of both performance and metacognition \cite{vaccaro2024combinations}, has been used in prior human-AI interaction research \cite{FERNANDES2026108779}, and serves as a benchmark for evaluating LLMs \cite{katz2024gpt}.

Across five outcomes (task performance, metacognitive accuracy, task load, trust, and usability), our results converge on a consistent dissociation: trace exposure shifts subjective appraisal of the interaction without bringing performance benefits. Summary traces preserved task performance at the no-trace baseline while significantly elevating trust and hedonic appeal  which provides a contrast in which the underlying LLM model was held constant (GPT-5). Under an open-weight reasoning model exposing verbose intermediate output (GPT-oss-20b), full traces additionally impaired performance relative to the \textit{Answer-only} baseline. Across all three conditions, participants substantially overestimated their performance, and no trace format supported calibrated self-evaluation, however, overestimation was largest under \textit{Full-trace}. A mediation analysis indicates that hedonic appeal, not trust, carries the indirect path to overestimation, consistent with a processing-fluency account of metacognitive misattribution.
Our contributions are threefold. First, we provide a \textit{controlled comparison of reasoning-trace formats} in terms of cognitive and metacognitive performance. We isolate verbosity as a design dimension and show that summarized traces preserve performance without improving calibration while elevating subjective appraisal of the interaction.
Second, we show that \textit{summaries preserve performance without improving calibration} while raising trust and hedonic appeal, and that under an open-weight verbose-trace regime, full traces additionally impair performance and produce the largest overestimation gap. This implies that calibration is unlikely to emerge from the trace itself, regardless of verbosity.
Third, we translate these findings into \textit{concrete design directions} for trace-based LLM interfaces: we argue that reasoning traces are best understood as user-facing interface artifacts rather than windows into model cognition. Metacognitive support must instead be scaffolded by interactions that elicit users' own reasoning before exposing the model's reasoning traces or answer.

\section{Related Work}

In this section, we focus on the role of metacognitive processes in human cognition and human–AI interaction. We first introduce central concepts of human metacognition and examine how mechanisms such as self-monitoring and evaluation influence problem-solving, learning, and decision-making. We then consider how these processes are affected when people collaborate with AI systems, highlighting both opportunities and risks for cognitive performance and user judgment. Finally, we introduce key concepts related to reasoning traces and AI explanations, which aim to increase transparency but may also introduce new challenges for calibration and usability.

\subsection{Metacognition and Human Reasoning}\label{sec:human_metacognition}
Metacognition refers to the processes by which people monitor, plan, and evaluate their own thinking \cite{fleming2024metacognition}. These processes are essential for complex problem-solving, learning, and optimizing behavior. Two primary constructs have been emphasized in psychological research: metacognitive accuracy and metacognitive sensitivity.

Metacognitive accuracy describes how closely individuals' self-evaluations align with their actual performance \cite{fleming2024metacognition, colombatto2023illusions}. High accuracy allows people to recognize their limitations and make informed adjustments, such as seeking more information, revising a strategy, or making a conservative choice. Accuracy can be further decomposed into bias (systematic over- or underestimation) and noise (random fluctuations) \cite{fiedler2019metacognition}.

Metacognitive sensitivity refers to the ability to discriminate between correct and incorrect judgments, often measured through confidence ratings after the decision \cite{fleming2024metacognition}. High metacognitive sensitivity entails that an individual's confidence ratings accurately reflect their performance: high confidence corresponds to correct judgments, and low confidence to incorrect ones.

Both constructs show substantial individual differences \cite{kelemen2000individual, ackerman2017meta} and are affected by external aids. People often misattribute externally generated information to themselves \cite{johnson_source_1993, Zindulka2026AIMemoryGap}, and reading fluent explanations inflates perceived understanding even when comprehension has not actually improved \cite{fisher2021harder}. These findings establish an underlying expectation that any system that provides articulate external reasoning may shift users' self-evaluations, regardless of whether it improves their actual performance.

\subsection{Metacognition in Human-AI Interaction}

Augmenting human intellect has long been a core theme in HCI, as highlighted by \citet{engelbart1962augmenting}. Research has emphasized the potential of AI for improving human performance \cite{10.1145/3411764.3445717, steyvers_bayesian_2022,zulfikar_memoro_2024, tankelevitch2025understanding}.
However, performance gains come with hidden risks related to how users perceive and rely on these systems. Users frequently hold inflated expectations of their performance with AI and fail to monitor their actual outcomes \cite{FERNANDES2026108779, kloft2023ai, villa2023placebo, kosch2023placebo, colombatto2025metacognition, von2025knowing}. Recent discussions also raise concerns about biases, skill loss, and deskilling, especially when generative AI improves short-term performance but undermines learning and long-term ability \cite{bastani2024generative, rafner2022deskilling, Kobiella2024, stadler_cognitive_2024}.

In AI-assisted tasks, the central limitation of human performance therefore appears to be primarily metacognitive, involving challenges in planning, monitoring, evaluating, and understanding AI interactions \cite{tankelevitch2023metacognitive}. A plausible explanation is that current AI interfaces are suboptimal at supporting metacognition \cite{ramesh2026metacognitive}. Design interventions that target this layer can reduce overreliance, enhance performance, and foster a sense of agency (e.g., tailoring AI interactions to users' metacognitive processes through prompt design, task decomposition, or natural-language uncertainty cues, e.g., ``I'm not sure, but...'') \cite{tankelevitch2023metacognitive, draxlerghost24, kim2024m}. 

Recent work has also begun to develop dedicated instruments for measuring metacognition in this setting. The \textit{Collaborative AI Metacognition Scale} captures the extent to which users actively plan, monitor, and evaluate their own thinking while working with AI. This treats metacognition not as a stable trait but as a process that interfaces can support \cite{Sidra03042026}. Consequently, designing human–AI systems that explicitly support metacognition is crucial to reduce deskilling while maintaining user agency and control. What remains unclear is how the format of an AI's reasoning output shapes these processes and whether the same explanation that supports monitoring in one form may inflate perceived understanding in another.

\subsection{Reasoning Traces and AI Explanations}
LLM-based AI assistants, such as ChatGPT, offer real-time guidance that, in principle, could help users align confidence with correctness. 
Currently, interfaces progressively expose reasoning traces, step-by-step justifications that resemble the model's ``chain-of-thought''. While AI traces promise transparency, prior work shows that explanations can sometimes be uninformative, or even ignored by users \cite{vasconcelos2023explanations, wang2021explanations}. They can also introduce new biases: full reasoning traces may increase perceived understanding without improving users' ability to evaluate correctness. As discussed in Section~\ref{sec:human_metacognition}, fluent external information inflates perceived understanding without a corresponding gain in judgment, and users can misattribute the AI's competence to themselves, inflating performance estimates while metacognitive accuracy falls~\cite{tankelevitch2023metacognitive}.

Studies on Explainable AI (XAI) report mixed effects. Traces can help users follow complex steps and improve task accuracy, but they may introduce new cognitive biases \cite{bertrand2022cognitive} and anchor users to the model's conclusion even when it is wrong \cite{10.1145/3397481.3450639}. Even placebic explanations, i.e., content-free justifications that convey no diagnostic information, can influence users' trust in a system, indicating that the presence of an explanation can shape user judgment independently of its informational content \cite{eiband2019impact}. Long, detailed explanations can also increase task load and time-on-task without proportional benefit, while concise justifications can be easier to use but risk omitting information relevant for evaluation.
In addition, reasoning traces can be misleading and are neither necessary nor sufficient for trustworthy interpretability \cite{barez2025chain}, as recent work shows that a model's ``chain-of-thought'' output and the internal computations that produce its answer can diverge. Thus, what the model verbalizes as ``reasoning'' may not faithfully represent how it actually arrived at its answer \cite{lindsey2025biology, 10.5555/3666122.3669397}.
\citet{vasconcelos2023explanations} also argue that there is a cognitive cost of engaging with an explanation that can affect overreliance, predicting that effortful explanations reduce inappropriate reliance on AI. Our Full-trace condition operationalizes a high-cost variant of this design and provides a direct test of whether this moderation generalizes to LLM reasoning traces.

These mixed effects highlight that while metacognition is critical to human problem-solving and decision-making, it remains underexplored in the context of AI-assisted reasoning and HAI. Current systems often improve performance without supporting calibration, and explanation formats such as reasoning traces may further complicate achieving this balance. Explanations vary widely in verbosity, ranging from no traces displayed to short summaries to full ``chain-of-thought'' outputs, yet the literature has not isolated this dimension or examined how it shapes users' metacognitive processes during AI-assisted reasoning tasks.

\section{Research Model}

Our study systematically investigates the benefits and drawbacks of reasoning traces for task performance, how they shape metacognition and calibration, and the usability and user-experience trade-offs users may encounter across explanation formats.

Note that we investigate traces from a user-facing perspective, that is, as a class of XAI output that users respond to regardless of its internal validity. Thus, the relevant question for our study is not whether a trace accurately reflects the model's computation, but whether and how it shapes the user's metacognitive processes. This framing is consistent with the XAI literature's distinction between proxy and faithful explanations \cite{barez2025chain, jacovi-goldberg-2020-towards, 10.5555/3666122.3669397}, and positions the faithfulness problem as a further design risk: Users may calibrate their reliance on outputs that are themselves unreliable representations of model ``thinking'', amplifying the metacognitive consequences.

We treat the content format of the trace as a single design dimension, ranging from no reasoning traces to summaries to full reasoning traces, and ask how variation along this dimension shapes performance, metacognition, trust, task load, and usability. 

Five preregistered research questions and corresponding hypotheses structure this contrast:

\begin{itemize}
    \item \textbf{RQ1 Metacognitive accuracy:} As fluent or verbose external information has been shown to inflate perceived understanding without improving judgment \cite{fisher2021harder, alter2009uniting}, we expect summaries of reasoning traces to support more accurate self-evaluation than full reasoning traces or no traces conditions. That is H1: \textit{Summary-trace}$>$\textit{Full-trace} $>$ \textit{Answer-only} in metacognitive accuracy.
    \item \textbf{RQ2 Task load:} Furthermore, as long explanations increase reading time and effort without proportional benefit \cite{vasconcelos2023explanations, bertrand2022cognitive}, we expect full reasoning traces to impose the highest task load. That is H2: \textit{Full-trace} $>$ \textit{Answer-only} and \textit{Full-trace} $>$ \textit{Summary-trace}; \textit{Summary-trace} $\gtrsim$ \textit{Answer-only} on the NASA-TLX.
    \item \textbf{RQ3 Trust and confidence:} Because fluent justifications inflate perceived understanding \cite{fisher2021harder}, and traces can shape user judgment independently of their content validity, triggering the processing-fluency heuristic \cite{alter_overcoming_2007}, the most verbose format should yield the highest confidence and trust. That is H3: \textit{Full-trace} $\geq $\textit{Summary-trace} $>$ \textit{Answer-only}.
    \item \textbf{RQ4 Objective performance:} As explanations can scaffold complex reasoning \cite{10.1145/3411764.3445717,tankelevitch2023metacognitive}, we predict that any trace will improve accuracy over an answer-only baseline. That is H4: \textit{Full-trace} $\geq$ \textit{Summary-trace} $>$ \textit{Answer-only}.
    \item \textbf{RQ5 Usability preference:} Finally, because concise justifications are typically easier to use while verbose ones risk overwhelming users \cite{MILLER20191}, we expect reasoning trace summaries to be preferred over full reasoning traces. That is H5: \textit{Answer-only} $\geq$ \textit{Summary-trace} $>$ \textit{Full-trace} on the SUS.
\end{itemize}

Taken together, these predictions imply that reasoning traces should yield more trust, more confidence, and more accurate task performance and, at least in summarized form, lead to better calibration, with task load as the main cost of full traces.

\section{Method}

In the following, we motivate and document our methodological choices when conducting our study. The research software and the analysis, along with all associated measures, can be found at [anonymized for review]. All data collected for the purpose of our paper and analysis scripts can be found at [anonymized for review].

\subsection{Study Design}
We build on the implementation introduced by \citet{FERNANDES2026108779}, which examined how AI assistance can improve performance while undermining metacognitive accuracy. We extended this approach to focus on LLM reasoning traces. 
The study uses a between-subjects design with three conditions: \textit{Answer-only}, \textit{Full-trace}, and \textit{Summary-trace}. The pre-registered sample is $N=570$ ($190$ per condition), based on an a priori power analysis targeting the smallest effect of interest. After exclusions, the final analyzed sample was $N=559$.%

\subsubsection{Models and Interface Rationale.}
Each problem appeared on the left-hand side of the screen, while an AI chat (using GPT-5 or gpt-oss-20b, depending on condition) was displayed on the right. 

We used OpenAI's ChatGPT due to its widespread adoption in cognitive performance tasks \cite{bastani2024generative,draxlerghost24,draxler2023gender} and because recent reasoning models allow access to a reasoning summary. Since GPT-5 provides summaries rather than full reasoning traces, we employed gpt-oss-20b for \textit{Full-trace} condition, as it openly exposes complete reasoning traces. GPT-5 was used in conditions \textit{Answer-only} and \textit{Summary-trace} (see Section~\ref{sec:task-description}).

We note that some chat interfaces, such as OpenAI's ChatGPT, typically present reasoning traces as a drop-down element that users can optionally expand. This structure was not included as a fourth condition in the current study as it would have brought another factor to the experimental design.

\subsubsection{Stimuli Control.}\label{sec:stimuli_control}
To avoid confounding trace format with model answer quality, items were pre-screened so that gpt-oss-20b and GPT-5 achieve the same performance on retained items (reducing model-ability confounds), while intentionally including a small number of items for which that shared answer was incorrect. The final set comprised ten items meeting these criteria. The models' performance on the ten retained LSAT tasks was $M=5.00$. Model identity and any correctness annotations were not disclosed to participants.

\subsection{Task Description}\label{sec:task-description}

\begin{figure*}[!htp]
    \centering
    \includegraphics[width=\linewidth]{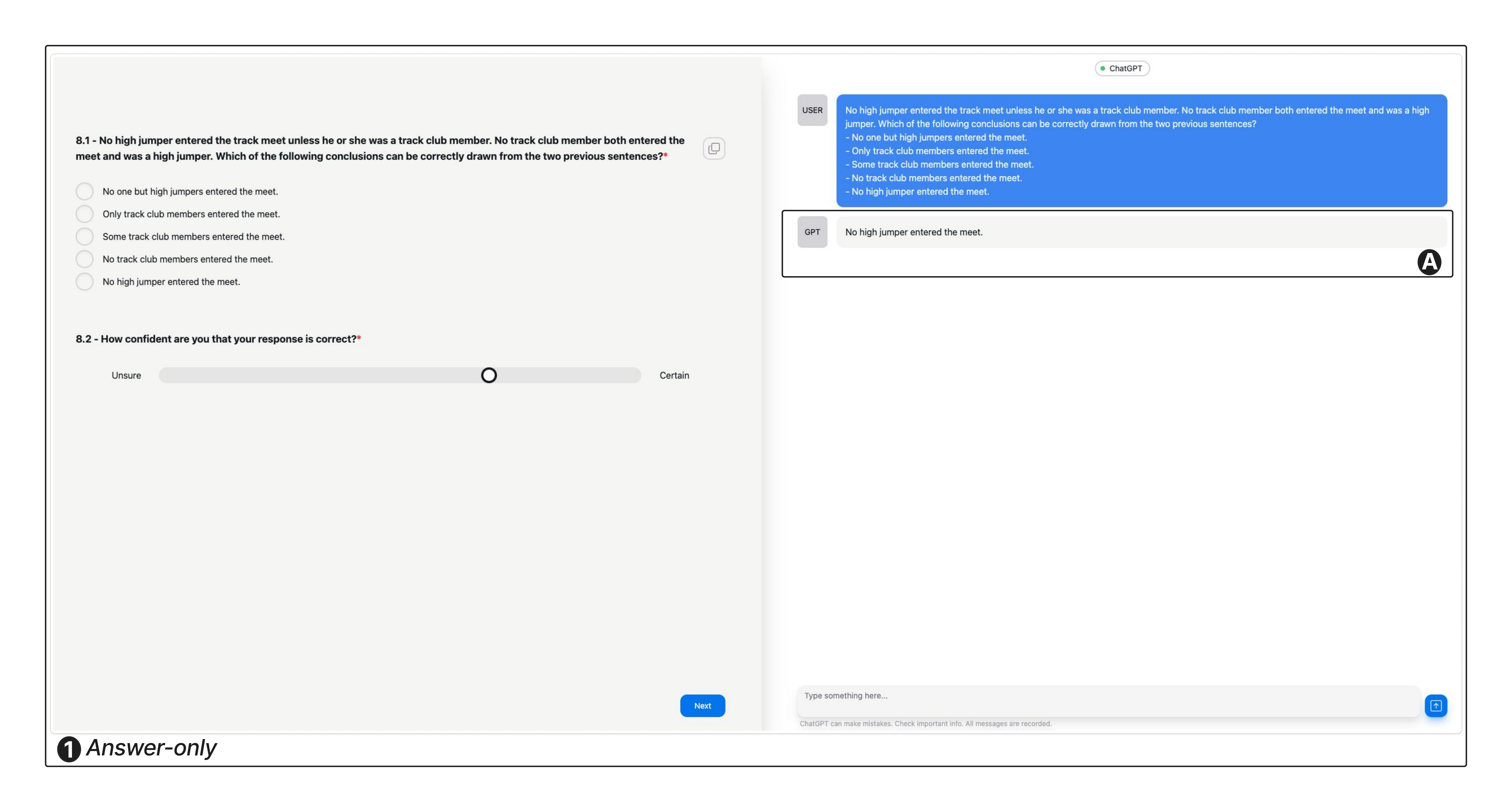}
    \caption{\textit{Answer-only} interface. The survey panel on the left presents the LSAT-reasoning problems, answer options, and the trial-level confidence slider. The chat panel on the right hosts the AI assistant. After the participant prompts the model, the assistant returns only the final answer (A).}
    \label{fig:answer_only}
\end{figure*}

\begin{figure*}[!htp]
    \centering
    \includegraphics[width=0.9\linewidth]{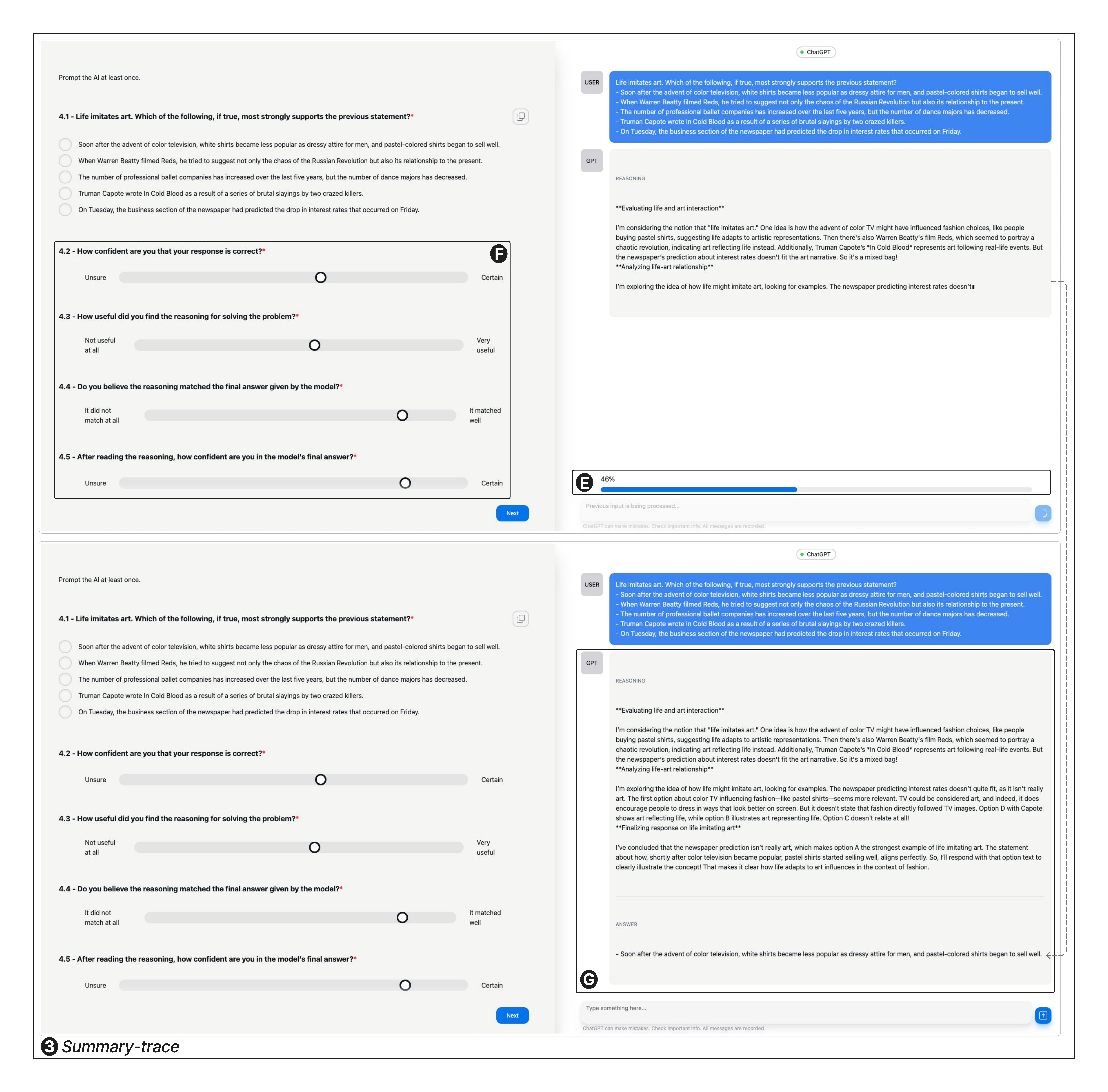}
    \caption{\textit{Summary-trace} interface. Left: the assistant streams a short reasoning summary into the chat with a progress bar (E) indicating generation. Right: once generation completes, the same response area shows the reasoning summary followed sequentially by the final answer (G), presented as a single message. The survey panel collects the same per-item ratings as in Full-trace (F).}
    \label{fig:summary_trace}
\end{figure*}

Participants completed ten LSAT logical reasoning items, presented in randomized order. The three experimental conditions were implemented as follows: In condition \textit{Answer-only} (\autoref{fig:answer_only}), participants saw only the model's final answer (\autoref{fig:answer_only}-\blackcircled{A}). In condition \textit{Full-trace} (\autoref{fig:full_trace}), the model's reasoning traces were displayed after each user prompt (\autoref{fig:full_trace}-\blackcircled{B}) while a progress bar (\autoref{fig:full_trace}-\blackcircled{E}) indicated that generation was underway. Once the reasoning trace was shown, a ``\textit{Final Answer}'' button (\autoref{fig:full_trace}-\blackcircled{C}) appeared, requiring participants to actively decide whether to view the model's final answer (\autoref{fig:full_trace}-\blackcircled{D}) or proceed with solving the reasoning problem without it. This workflow was deliberately designed to elicit a tentative answer from the participant before exposing them to the AI's final answer, thereby encouraging independent reasoning prior to model influence. In condition \textit{Summary-trace} (\autoref{fig:summary_trace}), the reasoning summary and the model's final answer were presented together, sequentially (\autoref{fig:summary_trace}-\blackcircled{G}).

After responding to each problem, on a trial-level, participants indicated their confidence by answering the question ``\textit{How confident are you that your response is correct?}'' using a 100-step slider ranging from unsure to certain. These ratings were used to compute indices of \textit{metacognitive sensitivity}, that is, the degree to which confidence discriminated correct from incorrect responses (see Section~\ref{sec:human_metacognition}).

In \textit{Full-trace} and \textit{Summary-trace}, as trial-level follow-up questions (\autoref{fig:full_trace}-\blackcircled{F}), participants also rated the perceived usefulness of the reasoning trace (\textit{``How useful did you find the reasoning for solving the problem?''}), whether it aligned with the model's final answer (\textit{``Do you believe the reasoning matched the final answer given by the model?''}) and how confident they were that the model's answer was correct based on the reasoning traces (\textit{``After reading the reasoning, how confident are you in the model's final answer?''}). These questions were answered using a 100-step slider. Participants were additionally provided with an open text box to record any comments about the reasoning traces or the model's final answer. As a measure of metacognitive bias, after completing all ten LSAT problems, participants provided a global estimate of their own performance by answering ``\textit{Using the AI, how many of the 10 logical reasoning problems do you think you solved correctly?}''.

To ensure engagement with the AI, participants were required to interact with the model at least once per problem before they could advance. Specifically, they needed to submit a prompt, wait for the AI's response to complete, enter their own answer, provide a confidence rating, and rate both the perceived usefulness and whether it aligned with the model's final answer. Beyond this minimum requirement, participants were permitted unlimited free-text interaction with the AI throughout the task. They were encouraged to use their own logical reasoning skills while optionally drawing on the AI's assistance to analyze and answer each question.

The LSAT reasoning task was selected for several key reasons. First, the LSAT is a widely recognized, real-world assessment used in high-stakes decision-making contexts such as law school admissions \cite{shultz2011predicting,wainer1995precision}. It has also been used as a benchmark for assessing large language model performance, making it particularly suitable for evaluating AI-assisted performance \cite{katz2024gpt}. An example of a logical reasoning question provided to participants was: Voltaire once said, ``Common sense is not so common.'' Which of the following most nearly parallels Voltaire's statement? (1) Truth serum cannot contain the truth. (2) God must have loved the common man --- he certainly made enough of them. (3) The common good is not necessarily best for everyone. (4) Jumbo shrimp may not actually be very big. (5) Good people may not necessarily have good sense.

\begin{figure*}[!htp]
    \centering
    \includegraphics[width=0.9\linewidth]{full_trace_no_fill.pdf}
    \caption{\textit{Full-trace} interface. Top: GPT streams the full reasoning trace into the chat (B) while a progress bar (E) indicates that generation is underway. Bottom: once the trace finishes, a Final Answer reveal button (C) appears; clicking it reveals the model's final answer (D). The survey panel additionally collects per-item ratings of trace usefulness, alignment of reasoning with the final answer, and confidence in the model's answer (F), which are collected only in trace conditions. The button design ensured that participants engaged with the trace itself before being exposed to the model's answer.}
    \label{fig:full_trace}
\end{figure*}

\subsection{Apparatus}

Extending the setup by Fernandes et al. \cite{FERNANDES2026108779}, we developed a custom web-based interface that displayed an AI chat window alongside a survey interface in a side-by-side layout. This system used the models and realized the interface concepts as described above.

The custom interface was designed to support detailed data collection. All user interactions were logged, enabling qualitative chat analysis. Survey responses and chat logs were automatically recorded and saved at the end of the study. To streamline prompting, we included a button that copied each problem and its answer options to the clipboard so participants could easily paste them into the chat.

The application was built with JavaScript and deployed via Netlify. The front end was connected to a Python proxy server hosted on Heroku, which forwarded chat messages to the OpenAI API and stored study data (survey responses and chat interactions) in a MongoDB database.

\subsection{Participants}
To explore individual differences in cognitive and metacognitive performance, we recruited a larger sample of 570 participants, with 190 participants per condition.
We powered for the smallest effect of interest. We computed the proportion of \textit{p}-values < .05 to determine the optimal sample size for sufficient statistical power (80\%). With this, we found that a sample size of 570 participants is optimal for reliably detecting differences between the upper and lower quartiles in metacognitive accuracy.

We recruited 570 English-fluent participants located in the USA through Prolific. We included an attention check, requiring participants to read a short description of the study and task. They then answered three multiple-choice questions, one about the topic (logical reasoning), another regarding which option to choose when solving the problems (the best one) and how they would be able to receive extra compensation.
We excluded six participants for erroneous responses (e.g., exceeding the number of possible correct answers when estimating performance), three for low completion times, and one for incorrect age disclosure. 
Upon the qualitative data analysis stage, we found that one participant did not follow the study instructions and was therefore removed from the dataset.

We further analyzed data from 559 participants (identified as female 253; identified as male 304, identified as non-binary 2; Age: $M$ = 40.77, $SD$ = 12.06). When asked to estimate their English fluency ("\textit{Estimate your fluency in the English language}"), 521 participants reported themselves as native English speakers, 37 as fully fluent and one as conversationally fluent. No participants preferred not to disclose their language proficiency. 39 participants in our sample reported their highest educational degree to be a doctoral degree, 134 a higher tertiary education degree (Master's level), 232 a lower tertiary education degree (Bachelor's level) and 95 an upper secondary school / high school. 
39 participants had taken the LSAT before; their performance was slightly lower, $M$ = 5.56, $SD$ = 1.59, compared to those who have not taken it, $M$ = 5.95, $SD$ = 1.40, thus they were not excluded from the sample. We collected informed consent from each participant before the study in accordance with the Declaration of Helsinki guidelines \cite{world2013world}. Each participant was compensated eight pounds per hour. In accordance with national [anonymized] guidelines, 
this study did not require ethics approval as it involved minimal risk to participants, with no intervention beyond standard practice and no collection of sensitive personal data. 

\subsection{Procedure}

Upon entering the study via Prolific, participants were redirected to our custom web application, where they were first presented with a description of the task and completed an attention check adapted to their assigned condition. In \textit{Answer-only}, the attention check consisted of two questions: ``\textit{What subject are you going to be answering questions about?}'' and ``\textit{Which answer choice should you select?}''. In \textit{Full-trace} and \textit{Summary-trace}, participants answered the same two questions plus an additional one, ``\textit{How can you earn extra compensation?}''. Extra compensation incentivized engagement with the task and the model: if a participant's estimated number of correct answers matched their actual number of correct answers, they were told they would receive an additional 0.40 pounds. All participants received the extra compensation regardless of their actual performance.

Participants then provided informed consent and completed a pre-task user expectations questionnaire. They were briefly introduced to the task and allowed to test the chat interface to familiarize themselves with its functionality before completing the main task as described in Section~\ref{sec:task-description}.

After completing all ten reasoning problems, participants filled out the expectations questionnaire again in the past tense, followed by a trust questionnaire, the NASA-TLX, the User Experience Questionnaire -- Short (UEQ-S), the System Usability Scale (SUS), and the Collaborative AI Metacognition Scale~\cite{Sidra03042026}. Finally, participants provided demographic information, including age, gender, occupation, education level, English proficiency, and whether they had previously taken the LSAT.

The study took, on average, 43 minutes to complete in \textit{Answer-only}, 51 minutes in \textit{Full-trace}, and 59 minutes in \textit{Summary-trace}. After finishing, participants were redirected back to Prolific to register their completion of the study.

\subsection{Analysis}
We analyzed the data in two stages. First, we conducted quantitative analyses to compare the three experimental conditions on task performance (achieved and perceived performance), metacognitive accuracy (that is, how closely participants' estimated performance aligns with their actual performance), perceived workload through the NASA-TLX questionnaire, trust in XAI context through a XAI Trust scale, user experience through the User Experience Questionnaire - Short (UEQ-S), usability through the System Usability Scale (SUS), and self-reported metacognitive engagement through the Collaborative AI Metacognition Scale \cite{Sidra03042026}. For an overview of the descriptive statistics, see Table~\ref{tab:overall_descriptives}.

Second, we conducted a qualitative thematic analysis of participants’ open-ended responses using an inductive thematic approach \cite{clarke2017thematic} to better understand how participants used the AI chatbot and how they perceived the reasoning traces. This analysis focused on the full reasoning traces condition (\textit{Full-trace}) and the summary of reasoning traces condition (\textit{Summary-trace}), where participants were exposed to reasoning traces. Responses were grouped into recurring interaction patterns such as guided reasoning, independent verification, and perceived burden of the reasoning traces. We report the frequency of each theme and use representative quotes to interpret the quantitative findings.

\section{Results}

\subsection{Metacognitive Accuracy}

\subsubsection{Objective Task Performance}\label{sec:achieved_performance}
\begin{figure}[!htp]
    \centering
    \includegraphics[width=0.8\linewidth]{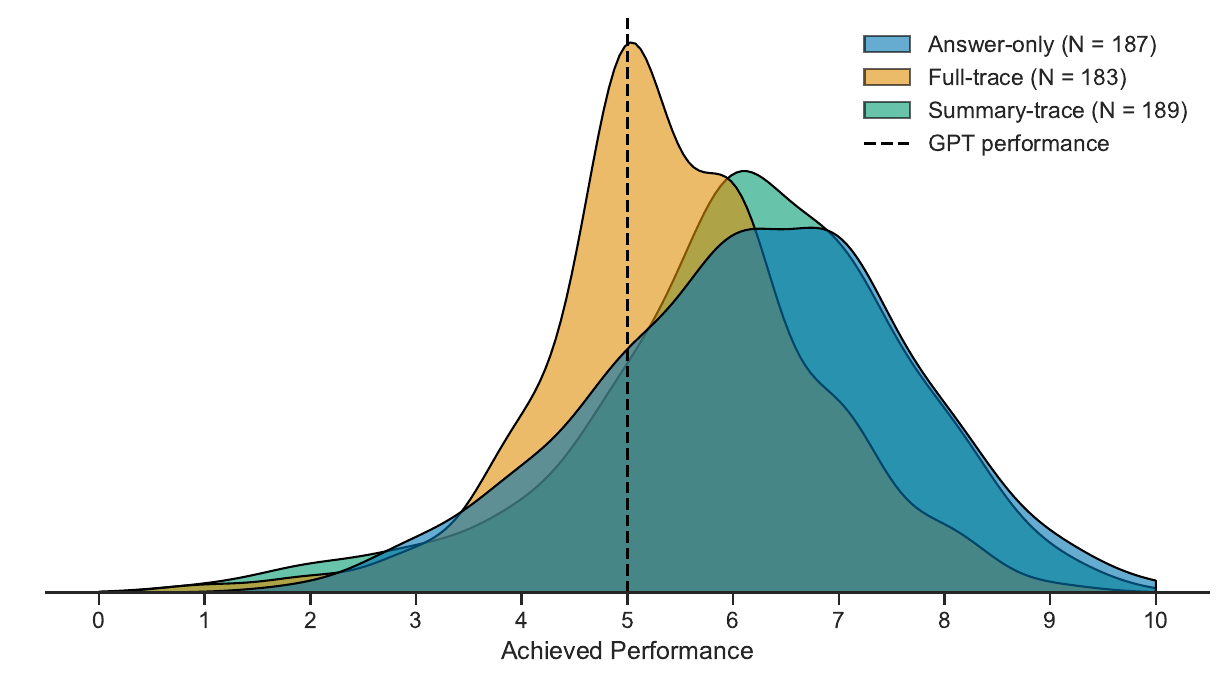}
    \caption{Achieved performance (0-10) across trace formats.  \textit{Full-trace} condition, $N=183$ (yellow curve); \textit{Summary-trace} condition, $N=189$ (green curve); \textit{Answer-only} condition, $N=187$ (blue curve). The performance of the models is represented by a dashed vertical line 
    (both GPT-5 and gpt-oss-20b achieved a score of 5 on the set of 10 LSAT questions).}
    \label{fig:performance_comparison}
\end{figure}

Achieved performance differed significantly across conditions (see Table~\ref{tab:overall_descriptives} and Figure ~\ref{fig:performance_comparison}), $F(2, 556) = 15.10$, $p < .001$, $\eta^2 = .052$. Participants in \textit{Answer-only} condition solved an average of 6.18 items ($SD = 1.46$), participants in \textit{Full-trace} condition solved an average of 5.46 items ($SD = 1.21$), and participants in \textit{Summary-trace} condition solved an average of 6.11 items ($SD = 1.46$).
\textit{Full-trace} participants performed significantly worse than \textit{Answer-only} ($t(368) = -5.15, p < .001, d = -0.54$) and \textit{Summary-trace} ($t(370) = -4.64, p < .001, d = -0.48$). \textit{Answer-only} and \textit{Summary-trace} conditions did not significantly differ ($t(374) = -0.47, p = .639, d = -0.05$).
Thus, full reasoning traces impaired task accuracy relative to both \textit{Answer-only} and \textit{Summary-trace} conditions, while the presence of a summary of reasoning traces preserved performance at the level of no reasoning traces (\textit{Answer-only} condition). 

As a benchmark comparison, we tested whether participants in each condition performed above the model-alone score $M$=5.00 (see Section \ref{sec:stimuli_control}). Participants in all three conditions scored significantly above the model-alone score: \textit{Answer-only} $t(186) = 11.11$, $p < .001$, $d = .81$; \textit{Full-trace}, $t(182) = 5.20$, $p < .001$, $d = .38$; and \textit{Summary-trace}, $t(188) = 10.44$, $p < .001$, $d = .76$. Figure~\ref{fig:performance_comparison} shows the distribution of achieved performance across the three conditions.

\subsubsection{Overestimation Bias}
\begin{figure*}[!htp]
    \centering
    \includegraphics[width=0.8\linewidth]{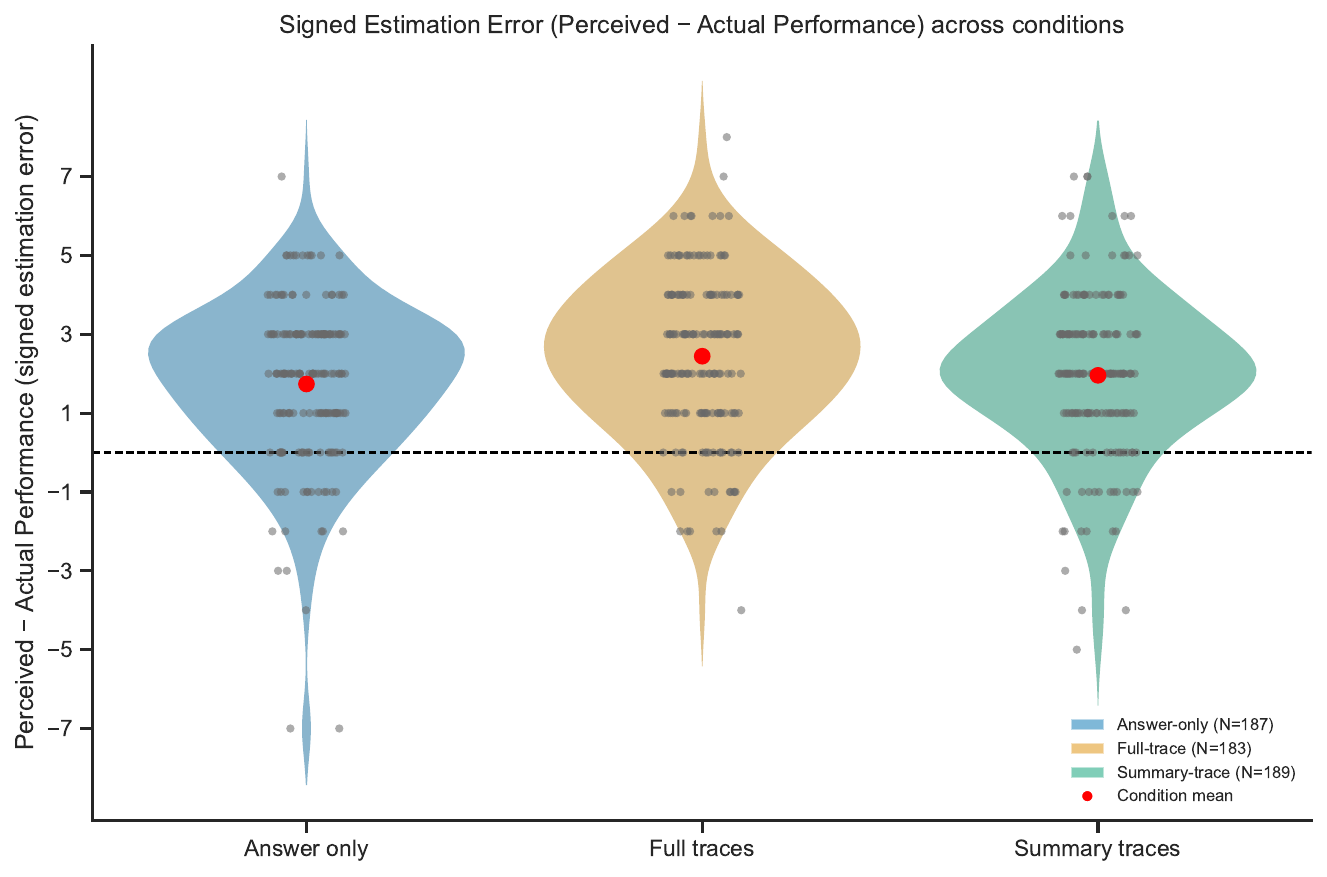}
    \caption{ Performance estimation bias (perceived -- actual performance) across conditions (\textit{Answer-only}, blue violin plot, N = 187), \textit{Full-trace}, yellow violin plot, N = 183, and \textit{Summary-trace}, green violin plot, N = 189). Red dots indicate condition means. The dashed horizontal line at zero marks perfect calibration. Positive values indicate overestimation and negative values underestimation.}
    \label{fig:violin_plot}
\end{figure*}

Perceived performance substantially exceeded actual performance in all conditions (see Table~\ref{tab:overall_descriptives} and Figure~\ref{fig:violin_plot}). In \textit{Answer-only} condition, participants estimated 7.92 items correct ($SD = 1.59$), while achieving 6.18 correct items, yielding a significant overestimation, $t(186) = 11.63$, $p < .001$, $d = .85$. In \textit{Full-trace} condition, participants estimated 7.91 items correct ($SD = 1.71$), while achieving 5.46 correct items, also showing significant overestimation, $t(182) = 16.35$, $p < .001$, $d = 1.21$. In \textit{Summary-trace} condition, participants estimated 8.07 items correct ($SD = 1.55$), while achieving 6.11 correct items, again showing significant overestimation, $t(188) = 13.28$, $p < .001$, $d = .97$.

Perceived performance did not differ significantly across conditions, $F(2, 556) = 0.58$, $p = .561$, $\eta^2 = .002$. However, the magnitude of overestimation differed significantly by condition, $F(2, 556) = 5.83$, $p = .003$, $\eta^2 = .021$. Participants in \textit{Full-trace} condition showed the largest overestimation bias ($M = 2.44$, $SD = 2.02$), followed by \textit{Summary-trace} ($M = 1.96$, $SD = 2.03$) and \textit{Answer-only} ($M = 1.74$, $SD = 2.04$). Tukey-adjusted post-hoc comparisons showed that overestimation was significantly larger in \textit{Full-trace} condition than in \textit{Answer-only} ($M_{\mathrm{diff}} = 0.70$, $p_{\mathrm{adj}} = .003$). \textit{Summary-trace} did not differ significantly from \textit{Answer-only} ($M_{\mathrm{diff}} = 0.22$, $p_{\mathrm{adj}} = .546$) or \textit{Full-trace} ($M_{\mathrm{diff}} = -0.48$, $p_{\mathrm{adj}} = .056$).

Together, these results indicate that participants across all conditions overestimated their performance, but this miscalibration was most pronounced in the full reasoning trace condition. Notably, \textit{Full-trace} participants performed worst objectively while perceiving their performance as comparable to participants in the other conditions, resulting in the largest overestimation gap. Summary traces did not significantly worsen miscalibration relative to the answer-only baseline, although participants' self-beliefs remained inflated.

\begin{table*}[!htp]
\centering
\caption{ANOVAs with descriptives by condition. Values are means with standard deviations in parentheses.}
\label{tab:overall_descriptives}
\begin{tabular}{lcccccccc}
\hline
DV                    & \textit{Answer-only}             & \textit{Full-trace}                 & \textit{Summary-trace} & F & df1 & df2 & p & \multicolumn{1}{l}{$\eta^2$} \\ \hline
Achieved  &    6.18 (1.46)      &      5.46 (1.21)     &       6.11 (1.46)           &       15.10           &              2        &      556                &      <.001***           &            .052         \\
Perceived &    7.92 (1.59)      &      7.91 (1.71)     &      8.07 (1.55)          &       0.58       &             2           &        556              &        .561        &           .002       \\
Overestimation &    1.74 (2.04)      &        2.44 (2.02)    &        1.96 (2.03)          &     5.83             &             2         &      556                &      .003**            &           .021           \\
NASA-TLX     & 43.85 (13.09)  & 45.19 (14.39)  & 45.58 (13.51)  &  .828 & 2 & 556 & .438 & .003  \\
TSQ                     & 3.61 (0.76)  & 3.80 (0.72)  & 3.83 (0.72)  & 4.99  & 2 & 556 & .007** & .018  \\
UEQ-S                     & 0.81 (0.92)  & 1.19 (0.79)  & 1.12 (0.83)  & 10.92  & 2 & 556 & <.001*** & .038  \\
SUS                     & 78.48 (15.34)  & 76.02 (19.32)  &  77.69 (17.15) & 0.962 & 2 & 556 & .383 & .003  \\ 

\hline
\end{tabular}

\raggedright \small \textit{Note}. Reported are descriptive statistics and one-way between-subjects ANOVAs with condition as the factor. The $p$ column refers to the omnibus ANOVA test, not pairwise comparisons. Pairwise condition differences were examined using Tukey-adjusted post-hoc comparisons and are reported in the text where relevant. Achieved performance, perceived performance, and overestimation are reported as number of items out of 10. Overestimation corresponds to perceived minus achieved performance. NASA-TLX and SUS are reported on 0--100 scales. TSQ is reported on a 1--5 scale. UEQ-S is reported on a $-3$ to $+3$ scale, where higher values indicate a more positive user experience. $\eta^2$ denotes effect size. Significance levels: *$p < .05$, **$p < .01$, ***$p < .001$.
\end{table*}

\subsection{Metacognitive Sensitivity}

To examine whether participants' self-assessments were related to their actual performance, we computed within-condition correlations between perceived and achieved performance. In all three conditions, correlations were low and non-significant: \textit{Answer-only}, $r(185) \approx .10$, $p = .166$; \textit{Full-trace}, $r(181) \approx .07$, $p = .319$; \textit{Summary-trace}, $r(187) \approx .10$, $p = .173$. This indicates that participants' performance estimates were weakly related to their actual scores, regardless of trace format. In other words, we found no evidence that any condition helped participants form more accurate self-assessments after completing all the tasks, as the format of reasoning traces did not improve the alignment between perceived and achieved performance. 

To test whether participants discriminated correct from incorrect responses within each condition, we ran paired-samples $t$-tests comparing $\overline{c_{+}}$ and
$\overline{c_{-}}$ (mean confidence on correct trials and mean confidence on incorrect trials, respectively). The effect was significant in every condition: \textit{Answer-only}, $t(185) = 6.27$, $p < .001$, $d_z = 0.46$, $M_{\Delta} = 0.065$; \textit{Full-trace}, $t(182) = 4.98$, $p < .001$, $d_z = 0.37$, $M_{\Delta} = 0.047$; \textit{Summary-trace}, $t(188) = 5.61$,
$p < .001$, $d_z = 0.41$, $M_{\Delta} = 0.062$. Across conditions, however, the size of the gap did not differ (\textit{Answer-only}: $M = 0.065$, $SD = 0.142$; \textit{Full-trace}: $M = 0.047$, $SD = 0.128$; \textit{Summary-trace}: $M = 0.062$, $SD = 0.151$), $F(2, 555) = 0.84$, $p = .432$, $\eta^2 = .003$.\footnote{For discrimination and AUC, one participant had to be removed due to scoring 10/10.\label{footnote_remove}}

To assess whether AI exposure changed how well participants' confidence ratings tracked their actual performance at the trial level, we computed four trial-level metacognitive indices from the ten LSAT items per participant. Correctness was coded as $1/0$, and the 0--100 confidence slider was rescaled to $[0,1]$. Table~\ref{tab:meta-indices} reports the descriptives and ANOVA summaries for all four indices of metacognitive sensitivity (for an introduction, see \citet{rahnev2025comprehensive}). We employed AUC, Brier Score, calibration and discrimination\footref{footnote_remove},\footnote{We do not report meta-d' or M-ratio as a reliable estimation, even under the HMeta-d hierarchical Bayesian estimator, which requires substantially more trials per participant than our 10-item battery.}. 
As a first intuitive sensitivity index, we computed each participant's
\emph{confidence gap}, $\Delta_{\text{disc}} = \overline{c_{+}} - \overline{c_{-}}$, defined as the difference between mean confidence on correct trials ($\overline{c_{+}}$) and mean confidence on incorrect trials ($\overline{c_{-}}$). Positive values indicate that confidence was higher when the response was right than when it was wrong and values near zero indicate insensitivity. 

\begin{table}[H]
\centering
\caption{One-way ANOVAs with descriptives by condition for the four
metacognitive indices.}
\label{tab:meta-indices}
\resizebox{\linewidth}{!}{
\begin{tabular}{lccccllll}
\hline DV                                & \textit{Answer-only} M(SD)  & \textit{Full-trace} M(SD)      & \textit{Summary-trace} M(SD)   & F     & df1 & df2 & $p$       & $\eta^2$ \\ \hline
Discrimination ($\Delta_{\text{disc}}$)  & 0.065 (0.142)   & 0.047 (0.128)   & 0.062 (0.151)   & 0.84  & 2   & 555 & .432      & .003 \\
AUROC2                                   & 0.586 (0.206)   & 0.562 (0.192)   & 0.587 (0.202)   & 0.97  & 2   & 555 & .380      & .003 \\
Calibration bias                         & 0.189 (0.179)   & 0.265 (0.183)   & 0.222 (0.178)   & 8.31  & 2   & 556 & $<$.001***& .029 \\
Brier score                              & 0.289 (0.111)   & 0.342 (0.117)   & 0.296 (0.111)   & 12.20 & 2   & 556 & $<$.001***& .042 \\
\hline
\end{tabular}
}

\raggedright \small \textit{Note}. Values are means with standard deviations in
parentheses. All indices are computed at the individual level (one
value per participant) on a $[0,1]$ scale. Stars denote uncorrected $p$-values (* $<.05$, ** $<.01$, *** $<.001$); 
\end{table}

As a second, distribution-free sensitivity index, we computed the non-parametric Type-2 ROC area, defined as the probability that a randomly drawn correct trial received a higher confidence rating than a randomly drawn incorrect trial, with ties contributing $0.5$. AUROC2 is bounded in $[0,1]$, equals $0.5$ at chance, and unlike
$\Delta_{\text{disc}}$ depends only on the rank-order of confidence, which makes it robust to wicked distributions. Mean AUROC2 was nearly identical across conditions
(\textit{Answer-only}: $M = 0.586$, $SD = 0.206$; \textit{Full-trace}: $M = 0.562$, $SD = 0.192$;
\textit{Summary-trace}: $M = 0.587$, $SD = 0.202$) and conditions did not differ, $F(2, 555) = 0.97$, $p = .380$, $\eta^2 = .003$.

Whereas the two indices above quantify \emph{sensitivity} (whether confidence \emph{discriminates} correct from incorrect trials), we
next examined \emph{calibration} (whether confidence \emph{matches} accuracy in absolute terms). For each participant, we computed
$\mathrm{calibration} = \overline{c} - \mathrm{accuracy}$, where $\overline{c}$ is mean confidence and $\mathrm{accuracy}$ is the proportion of items answered correctly. Positive values indicate overconfidence, negative values indicate underconfidence. Calibration differed between conditions, $F(2, 556) = 8.31$, $p < .001$, $\eta^2 = .029$.
All conditions were overconfident on average, but the \textit{Full-trace} condition ($M = 0.265$, $SD = 0.183$) showed greater overconfidence
than the \textit{Answer-only }baseline ($M = 0.189$, $SD = 0.179$; Tukey HSD, $p < .001$); the \textit{Summary-trace} condition ($M = 0.222$, $SD = 0.178$) fell between the
two and did not differ reliably from either (vs.\ \textit{Answer-only}, $p = .18$; vs.\ \textit{Full-trace}, $p = .056$).

As an integrated measure that combines calibration and resolution, we computed the Brier score, $\mathrm{BS} = \tfrac{1}{n}\sum_{i}(c_i - y_i)^2$, i.e.\ the mean squared error of confidence treated as a probabilistic forecast of correctness $y_i \in \{0,1\}$. It rewards forecasts that are both well-calibrated and discriminating, and lower values indicate forecasts that are simultaneously well-calibrated and informative; a participant who reports a confidence of 50 on every trial attains a Brier score of $0.25$. Conditions differed reliably, $F(2, 556) = 12.20$, $p < .001$, $\eta^2 = .042$. The \textit{Full-trace} condition produced the worst (highest) Brier scores ($M = 0.342$, $SD = 0.117$) and was reliably worse than both \textit{Answer-only} ($M = 0.289$, $SD = 0.111$; Tukey HSD, $p < .001$) and \textit{Summary-trace} ($M = 0.296$, $SD = 0.111$; $p < .001$); \textit{Answer-only} and \textit{Summary-trace} did not differ ($p = .80$). Together, these results indicate that participants' metacognitive sensitivity was relatively weakened in the \textit{Full-trace} condition.

\subsubsection{Collaborative AI Metacognition Scale (CAIMS)}
We use CAIMS as a self-report index of users' disposition to engage metacognitively with AI.
Metacognitive engagement during AI-assisted reasoning was assessed using the Collaborative AI Metacognition Scale \cite{Sidra03042026}, which measures Planning, Monitoring, and Evaluation of one's thinking while working with AI (all items rated on a 1--5 scale; higher scores indicating a disposition for greater metacognitive engagement in AI Interaction).

Self-reported metacognitive collaboration on the CAIMS was high across the board and increased monotonically from \textit{Answer-only} through \textit{Full-trace} to \textit{Summary-trace} on every
subscale (see Table~\ref{tab:caims-means}); the one-way ANOVA reached
significance only for evaluation, $F(2, 556) = 3.02$, $p = .049$,
$\eta^2 = .011$ (\textit{Summary-trace} $>$ \textit{Answer-only}, Tukey $p = .038$). The correlational picture, however, ran counter to what a genuine
metacognitive-monitoring account would predict: pooling across all 559 participants (see Table~\ref{tab:caims-overall}), every CAIMS scale
predicted \emph{larger} signed overestimation bias and \emph{larger} absolute metacognitive error (all $r \approx .12$--$.18$,
$p \leq .005$, Holm-significant).

Extending the analysis to the trial-level indices of metacognitive sensitivity reinforces the pattern: CAIMS
positively predicted mean confidence (planning $r = .27$, monitoring $r = .26$, evaluation $r = .33$, total $r = .33$), trial-level calibration error (all $r \approx .25$--$.30$), and Brier scores (all $r \approx .21$--$.26$; all of these Holm-significant at
$p < .001$), and showed small \emph{negative} associations with the two rank-based sensitivity indices (confidence gap and AUROC2; planning, evaluation, and total Holm-significant for both indices, $r$'s between $-.13$ and $-.10$; monitoring not surviving correction). Self-reported metacognitive literacy, therefore, tracks the \emph{level} of inflated confidence and the calibration error it
produces while, if anything, being inversely related to the \emph{informativeness} of confidence.

Crucially, the effect is not specific to the trace conditions: in
\textit{Answer-only} evaluation still predicted absolute miscalibration, $r = .23$, $p = .002$,
Holm $p = .006$, so the correlation cannot be reduced to the manipulation. The signed bias correlations were nevertheless
strongest in the \textit{Summary-trace} condition (planning $r = .25$, monitoring $r = .28$, evaluation $r = .20$, total $r = .28$; all $p \leq .006$),
accompanied by a slight performance drop (planning $\rightarrow$ score $r = -.18$, $p = .014$), indicating that \textit{Summary-trace}-condition participants who reported more metacognitive interaction literacy in terms of planning skewed specifically toward overestimation. This largely replicates \citet{FERNANDES2026108779}.

\begin{table*}[!htp]
\centering
\caption{Overall (condition-agnostic) Pearson correlations between
CAIMS subscales and the three outcome variables.} 
\label{tab:caims-overall}
\begin{tabular}{lccc}
\hline Subscale     & Score                                & Calibration bias ($\Delta$)             & Absolute accuracy ($|\Delta|$)        \\ \hline
Planning             & $-.09^{*\dagger}$ $[-0.17,\,-0.00]$  & $.15^{***}$ $[0.07,\,0.23]$            & $.12^{**}$ $[0.04,\,0.20]$            \\
Monitoring           & $-.07$ $[-0.16,\,0.01]$              & $.15^{***}$ $[0.07,\,0.23]$            & $.15^{***}$ $[0.06,\,0.23]$           \\
Evaluation           & $-.06$ $[-0.14,\,0.03]$              & $.16^{***}$ $[0.08,\,0.24]$            & $.16^{***}$ $[0.08,\,0.24]$           \\
Total                & $-.09^{*\dagger}$ $[-0.17,\,-0.00]$  & $.18^{***}$ $[0.09,\,0.25]$            & $.16^{***}$ $[0.08,\,0.24]$           \\
\hline
\end{tabular}

\raggedright \small \textit{Note}. Values are $r$ with 95\,\% CI; $N = 559$. Stars denote uncorrected $p$-values (* $<.05$, ** $<.01$, *** $<.001$); $\dagger$ marks effects that do not survive Holm correction across the 16 tests within each outcome.

\end{table*}

\begin{table*}[!htp]
\centering
\caption{CAIMS subscale and total descriptives by condition. }
\label{tab:caims-means}
\resizebox{\linewidth}{!}{
\begin{tabular}{lccccrrrr}
\hline Subscale     & Overall M(SD) & \textit{Answer-only} M(SD) & \textit{Full-trace} M(SD)   & \textit{Summary-trace} M(SD) & $F$  & df1 & df2 & $p$ \\ \hline
Planning             & 4.02 (0.70)   & 3.95 (0.72)    & 4.00 (0.67)  & 4.09 (0.72)   & 1.94 & 2   & 556 & .145   \\
Monitoring           & 4.04 (0.66)   & 3.97 (0.71)    & 4.04 (0.65)  & 4.10 (0.61)   & 1.65 & 2   & 556 & .193   \\
Evaluation           & 4.22 (0.64)   & 4.14 (0.66)    & 4.22 (0.63)  & 4.30 (0.63)   & 3.02 & 2   & 556 & .049*  \\
Total                & 4.08 (0.59)   & 4.01 (0.60)    & 4.08 (0.58)  & 4.15 (0.57)   & 2.71 & 2   & 556 & .067   \\
\hline
\end{tabular}
}

\raggedright \small \textit{Note}. Values
are means with standard deviations in parentheses on the 1--5 metric;
$F$ and $p$ are from one-way ANOVAs comparing the three
conditions.

\end{table*}

\subsection{Task Load}

We next examined perceived workload using the NASA-TLX. Overall task load did not differ significantly across conditions (\textit{Answer-only}: $M=43.85,SD=13.09$; \textit{Full-trace}: $M=45.19,SD=14.39$; \textit{Summary-trace}: $M=45.58,SD=13.51$; $F(2, 556) = 0.83$, $p = .438$, $\eta^2 = .003$. This suggests that exposure to reasoning traces, whether \textit{Full-trace} or \textit{Summary-trace}, did not impose additional task load relative to receiving only a final answer. Descriptive statistics for the overall NASA-TLX score and its six subscales are shown in Table~\ref{tab:nasa}.

Trust in the AI differed significantly across conditions $(F(2,556) = 5.00, p = .007,\eta^2 = .018)$. In all three conditions, trust ratings were significantly above the scale midpoint ($\mu = 3$): \textit{Answer-only} ($M = 3.61, SD = 0.76; t(186) = 10.91, p < .001, d = 0.80$), \textit{Full-trace} ($M = 3.80, SD = 0.72; t(182) = 15.15, p < .001, d = 1.12$), and \textit{Summary-trace} ($M = 3.83, SD = 0.72; t(188) = 15.76, p < .001, d = 1.15$). Tukey-adjusted post-hoc comparisons revealed that participants in the \textit{Answer-only }condition reported significantly lower trust than those in both the \textit{Full-trace} condition ($p_{adj} = .030$) and the \textit{Summary-trace} condition ($p_{adj} = .011$). No significant difference was found between the \textit{Full-trace} and \textit{Summary-trace} conditions ($p_{adj} = .941$).

\subsubsection{Follow-Up Ratings}
After each problem, participants rated their confidence in their own answer on a 0--100 slider. This trial-level confidence measure differs from the global performance estimate reported after task completion (used to compute overestimation above, see section \ref{sec:achieved_performance}). In Conditions \textit{Full-trace} and \textit{Summary-trace}, participants additionally rated the perceived usefulness of the reasoning traces, whether the reasoning matched the model's final answer, and their confidence in the model's answer (descriptive statistics reported in Table~\ref{tab:follow_ups}). Confidence in one's own answer did not differ significantly across conditions ($F(2,556) = 2.05, p = .129, \eta^2 = .007$). Among conditions exposed to reasoning traces, participants did not differ significantly in perceived usefulness of the reasoning ($F(1,370) = 3.05, p = .081, \eta^2 = .008$) or confidence in the model's answer ($F(1,370) = 0.63, p = .428, \eta^2= .002$). However, participants in the \textit{Full-trace} condition rated the reasoning as matching the model's final answer more closely than those in the \textit{Summary-trace} condition $(F(1,370) = 4.72, p = .030, \eta^2= .013$; \textit{Full-trace}: $M = 80.64, SD = 14.48; \textit{Summary-trace}: M = 76.87, SD = 18.64)$. This suggests that while full traces were perceived as more internally coherent, this did not translate into higher usefulness ratings or greater confidence in the model, which is consistent with \cite{von2025knowing}.

\begin{table*}[!htp]
\centering

\caption{ANOVAs with descriptives by condition for follow-up questions after each trial. Values are means with standard deviations in parentheses.}
\label{tab:follow_ups}
\begin{tabular}{lccccllll}
\hline DV             & \textit{Answer-only} & \textit{Full-trace}       & \textit{Summary-trace}    & F    & df1 & df2 & $p$     & $\eta^2$ \\ \hline
Confidence (own answer)   & 80.68 (12.39)  & 81.13 (15.03)  & 83.29 (12.42)  & 2.05 & 2  & 556 & .129  & .007     \\
Usefulness of reasoning   & --  & 78.56 (15.95)   & 75.13 (18.73)   & 3.05 & 1   & 370  & .081     &  .008  \\
Reasoning matched answer  & --  & 80.64 (14.48)  & 76.87 (18.64)  & 4.72 & 1   & 370  & .03*   & .013   \\
Confidence in model answer    & --   & 77.89 (16.23)   & 76.53 (16.88)   & .629 & 1   & 370  &  .428 & .002      \\
\hline
\end{tabular}

\raggedright \small  \textit{Note}: Reported are one-way ANOVAs with condition as the factor. For Confidence (in own answer), all three conditions (\textit{Answer-only}, \textit{Full-trace}, \textit{Summary-trace}) were included. For the other follow-ups, only \textit{Full-trace} vs. \textit{Summary-trace} conditions were compared (\textit{Answer-only }did not include these questions). Means are on a 0--100 scale. $\eta^2$ denotes effect size. Significance levels: *$p < .05$, **$p < .01$, ***$p < .001$.
\end{table*}

\subsection{System Usability}

\subsubsection{System Usability Scale} System usability was measured using the SUS questionnaire \cite{brooke1996sus,bangor2008empirical}, which did not differ significantly across conditions, $F(2, 556) = 0.96$, $p = .383$, $\eta^2 = .003$. Mean SUS scores were high across all conditions: \textit{Answer-only} ($M = 78.48$, $SD = 15.34$), \textit{Full-trace} ($M = 76.02$, $SD = 19.32$), and \textit{Summary-trace} ($M = 77.69$, $SD = 17.15$). Tukey-adjusted post-hoc comparisons showed no significant pairwise differences between conditions, all $p_{\mathrm{adj}} \geq .363$.

\subsubsection{User Experience}
We next examined participants' subjective user experience using the UEQ-S \cite{alberola2018creation}. The UEQ-S provides separate scores for Pragmatic Quality, reflecting task-oriented qualities such as clarity, efficiency, and ease of use, and Hedonic Quality, reflecting qualities such as interest, excitement, and novelty. Scores range from $-3$ to $+3$, with higher values indicating a more positive user experience. Descriptive statistics are reported in Table~\ref{tab:ueqs}.

\begin{table*}[!htp]
\centering
\caption{Pragmatic, Hedonic and Overall scales per condition for the UEQ-S questionnaire, M(SD)}
\begin{tabular}{lcccccccc}
\hline
                  & \textit{Answer-only}    & \textit{Full-trace}   & \textit{Summary-trace}   & F & df1 & df2 & $p$     & $\eta^2$ \\ \hline
Pragmatic Quality & 0.89 (0.66) & 1.07 (0.60) & 0.94 (0.71) & 4.01 & 2& 556 & .019* & .014 \\
Hedonic Quality   & 0.73 (1.39) & 1.32 (1.16) & 1.31 (1.21) & 13.31 & 2& 556& <.001*** & .046 \\
UEQ-S Overall     & 0.81 (0.92) & 1.19 (0.79) & 1.12 (0.83)& 10.92 &2&556&<.001*** & .038 \\  
\hline
\end{tabular}

\raggedright \small \textit{Note}: Reported are one-way ANOVAs with condition as the factor. All three conditions (\textit{Answer-only}, \textit{Full-trace}, \textit{Summary-trace}) were included. Means and Standard Deviations are reported on the UEQ-S scale, ranging from $-3$ to $+3$, where higher values indicate a more positive user experience. $\eta^2$ denotes effect size.  Significance levels: *$p < .05$, **$p < .01$, ***$p < .001$.
\label{tab:ueqs}
\end{table*}

Condition had a significant effect on all three UEQ-S measures. For Pragmatic Quality, Tukey-adjusted post-hoc comparisons showed that \textit{Full-trace} was rated significantly higher than \textit{Answer-only} ($p_{adj} = .017$), whereas \textit{Summary-trace} did not differ significantly from either \textit{Answer-only} ($p_{adj} = .728$) or \textit{Full-trace} ($p_{adj} = .114$). For Hedonic Quality, both \textit{Full-trace} and \textit{Summary-trace} were rated significantly higher than \textit{Answer-only} (both $p_{adj} < .001$), with no significant difference between \textit{Full-trace} and \textit{Summary-trace} ($p_{adj} = .999$). The same pattern was observed for UEQ-S. Both \textit{Full-trace} ($p_{adj} < .001$) and \textit{Summary-trace} ($p_{adj} = .001$) were rated significantly higher than \textit{Answer-only}, while \textit{Full-trace} and \textit{Summary-trace} did not differ significantly ($p_{adj} = .703$).
Thus, reasoning traces improved subjective user experience relative to the \textit{Answer-only}, particularly for hedonic quality. However, \textit{Full-trace} and \textit{Summary-trace} were experienced similarly overall, suggesting that the more verbose full traces did not provide an additional subjective experience benefit over summarized traces.

\subsection{Mechanism: Does the trace-induced trust gain explain miscalibration?}
\label{sec:mediation}
The metacognitive sensitivity analyses above show that trace exposure, particularly exposure to full reasoning traces, inflates participants' confidence in their own performance, but they leave the mechanism underspecified.
One possibility is that the trace-rich interaction shifts participants' subjective appraisal of the AI (perceived trustworthiness, perceived engagement), and that they then use this appraisal as evidence of their joint competence. Because we collected both TSQ trust ratings and UEQ hedonic appeal in the same session, this account is testable within the present sample.
To probe the mechanism behind the condition effect, we fit a parallel multiple-mediator model in \texttt{lavaan} with condition as a three-level dummy-coded predictor (\textit{Answer-only} reference), trust (TSQoverall), and UEQ hedonic as parallel mediators, and overestimation bias ($\Delta = \mathrm{Estimate} - \mathrm{Score}$) as the outcome; all variables were $z$-standardized and indirect effects were tested with $5{,}000$ percentile bootstrap samples ($N = 559$).

Both trace conditions raised trust ($a_{1,\text{full}} = 0.26$, $p = .011$; $a_{1,\text{summary}} = 0.30$, $p = .005$) and hedonic appeal ($a_{2,\text{full}} = 0.45$; $a_{2,\text{summary}} = 0.45$; both $p < .001$), but only hedonic appeal predicted overestimation once trust was partialled out ($b_{2} = 0.15$, $p = .007$; $b_{1} = 0.09$, $p = .13$), so the indirect path through hedonic was significant for both conditions (\textit{Full-trace}: $\hat{\beta} = 0.067$, 95\,\% CI $[0.018, 0.129]$; \textit{Summary-trace}: $0.066$, $[0.018, 0.129]$) whereas the trust path was not. Single-mediator sensitivity models showed that trust \emph{does} carry an indirect effect on its own (\textit{Full-trace}: $0.047$, $[0.008, 0.098]$), but this effect is absorbed by hedonic when both are entered jointly: trust and hedonic correlate $r = .65$ ($R^{2} = .42$, tolerance $= .58$, VIF $= 1.72$ for trust and $1.77$ for hedonic in the $b$-path), so when entered together the trust slope shrinks by $51\%$ (solo $b = 0.18 \to$ joint $b = 0.09$) while the hedonic slope shrinks by only $28\%$ ($0.20 \to 0.15$). The two mediators jointly explained variance in overestimation over and above condition, $F(2, 554) = 13.01$, $p < .001$, but this shared variance is captured almost entirely by hedonic appeal, indicating that UEQ hedonic provides a more comprehensive proxy for the relevant subjective appraisal. The \textit{Full-trace} condition retained a direct effect ($c'_{f} = 0.25$, $p = .014$; partial mediation), while the \textit{Summary-trace} condition was fully mediated ($c'_{s} = 0.02$, $p = .89$).

The AI interaction inflates participants' subjective appraisal of the exchange, which they then use as evidence of their own performance. Once the variance shared by trust and hedonic is partitioned, only the unique-to-hedonic variance predicts overestimation, suggesting the active ingredient is closer to a fluency/affect signal than to calibrated trust, consistent with processing-fluency accounts of metacognitive misattribution \cite{alter2009uniting}.

\subsection{Qualitative Analysis}

To understand how participants engaged with reasoning traces (\textit{Full-trace} and \textit{Summary-trace} conditions), we conducted an inductive thematic analysis \cite{clarke2017thematic} of participants' open-ended responses to ``\textit{Please describe how you used the AI Chatbot}'' and ``\textit{Please share any comments about the reasoning or the answer}''. Across both conditions, four recurring interaction patterns emerged: cognitive offloading, guided reasoning, calibrated comparison, and trace burden. We report the frequency of each pattern and use representative quotes to interpret the quantitative findings\footnote{ C2 and C3 prefixes on participant IDs correspond to the \textit{Full-trace} and \textit{Summary-trace} conditions, respectively.}.

\subsubsection{Full-trace Condition}

In the \textit{Full-trace} condition, participants saw the model's reasoning trace before it revealed its final answer. Participants' descriptions referred almost exclusively to the reasoning content, and the reveal-answer button itself was rarely mentioned. We therefore interpret the patterns as responses to the traces.

Four main patterns emerged (see Table~\ref{tab:qualitative_C2}).
The largest group (Group 1, $35.3\%$) used the model as primary source of answers, asking it for answers or following its suggestions with limited independent evaluation. For example, participant C2-P122 wrote that the chatbot was ``\textit{spot on}'' and that they ``\textit{just followed exactly what it suggested}''. Similarly, C2-P130 described pasting the question and answer options and having the model ``\textit{choose the best answer}''. This pattern suggests that full traces led to delegation rather than support evaluation and functioning as a transparency mechanism. Instead, the reasoning seemed to act like the placebic explanations described by \citet{eiband2019impact}, where the presence of fluent justifications increased trust independently of whether participants engaged with its content. This interpretation is consistent with participants' trust ratings in the \textit{Full-trace} condition despite its lower task accuracy.

A second large group (Group 2, $57/183, 31.1\%$) used the trace as a guided reasoning aid, asking follow-up questions, using the model to break down problems, or relying on the trace to choose among options. Participant C2-P22 wrote that the model ``\textit{guided me step by step},'' and ``\textit{increased my confidence in selecting the right answer}.''
This pattern is the qualitative counterpart to the illusion of understanding documented by \citet{fisher2021harder}, where fluent, well-organized reasoning traces gave participants a sense that they had grasped the problem, even when their task performance did not improve. The same mechanism may explain the largest overestimation observed in the \textit{Full-trace} condition, where participants felt they had reasoned through the problem, when in practice, they simply trusted the model reasoning.

A third smaller group ($38/183, 20.8\%$) used the model for calibrated comparison. Participants first formed their own answer, then used the model's reasoning to confirm or challenge it. C2-P34 reported that while the AI was reasoning, they ``reasoned through the problem myself,'' and then checked the model’s logic against their own.
This pattern constitutes genuine human–AI synergy \cite{vaccaro2024combinations} in which the user retains the cognitive work and uses the trace as a point of comparison rather than a replacement for their own reasoning.

Finally, a notable subset (Group 4, $19/183, 10.4\%$) experienced the trace as a burden, being too long, confusing, or something they actively shortened. C2-P53 wrote that the responses were ``\textit{so incredibly long}'' that reading them thoroughly would take too much time. Participant C2-P138 reported asking the model to ``\textit{dumb it down}'' and ``\textit{summarize it to be shorter}'' when the reasoning was confusing. These comments highlight the problem in XAI literature that the same verbosity that promises transparency can become an obstacle, with several participants reproducing the \textit{Summary-trace} condition on their own initiative.

\subsubsection{Summary-trace Condition}

In the \textit{Summary-trace} condition, participants saw a short summary of the model's reasoning together with its final answer. When analyzing the qualitative data, three patterns accounted for 95\% of responses (see Table~\ref{tab:qualitative_C3}).

The largest group (39.2\%) used the traces as a reasoning and decision-support aid. These participants described reading the summarized reasoning, using it to evaluate options, or asking follow-up questions. C3-P64 wrote that the chatbot ``\textit{helped break down each question}'' and clarified relationships between statements and assumptions. The summary appeared to provide enough rationale to support reasoning without imposing the costs of a full trace, consistent with the \textit{Full-trace} burden pattern.

A second group (28.6\%) used the summary for calibrated comparison. These participants first formed their own answer, then compared it to the model's. C3-P58 provides a particularly clear example: ``\textit{(...)When I disagreed with the model, I checked its reasoning against my own. If I was not persuaded then I would stick with my answer}''. The proportion of calibrated comparison users was similar across the \textit{Full-trace} and \textit{Summary-trace} conditions (20.8\% vs. 28.6\%), suggesting that whether participants adopt this stance reflects the user as much as the interface.

A third group (27.5\%) used the model as their primary source of answers, copying questions into it and selecting the answer based on its output. C3-P127 stated, ``\textit{On most answers I solely relied on the AI Chatbot. I copy/pasted the question and answers and allowed it to give me the answer},'' Notably, this offloading rate was lower in the \textit{Summary-trace} condition than in \textit{Full-trace} (27.5\% vs. 35.3\%). Although both rates are substantial, the difference may suggest that a full trace may lead participants to delegate more strongly than a summary, perhaps due to its apparent thoroughness \cite{10.1145/3397481.3450639, eiband2019impact}, whereas a summary's brevity encourages users to keep reasoning alongside the model.

A small final group provided unclear or inconclusive descriptions of use (Group 5: $9/189, 4.7\%$). These responses often expressed general impressions.  

Overall, the \textit{Summary-trace} qualitative data suggest that many participants used the short reasoning summary to evaluate options or guide their decisions, and a substantial subset used it for calibrated comparison against their own reasoning. This pattern supports the idea that summaries may preserve some benefits of reasoning transparency while reducing the usability costs associated with full reasoning traces. At the same time, the low-interaction group shows that summaries do not eliminate cognitive offloading, as some participants still used the model primarily as a quick source of answers. 

\section{Discussion}

The five preregistered predictions divide into a single confirmation and four divergences, revealing an informative pattern. Contrary to H4, participants who were exposed to \textit{Full-trace} performed significantly worse and overestimated their performance to a greater degree than those who were not exposed to any trace. Qualitative patterns indicate that this is mediated by offloading and guided reasoning rather than by genuine critical thinking, as evidenced by the model traces. Traces acted as substitutes rather than supports for user reasoning. \textit{Summary-trace} preserved task performance at the no-trace baseline and led to lower overestimation levels, but no condition produced well-calibrated self-evaluation, providing no support for H1. Trace exposure inflated bias and left sensitivity unchanged, indicating that reasoning content does not feed the monitoring loop, which aligns with the notion that explanations are rarely used in HAI decision-making \cite{wang2021explanations}. H2 and H5 were also not supported, with null effects for task load and usability. H3, which predicted higher trust and confidence under trace exposure, was the only hypothesis the data partially supported: both trace conditions raised TSQ trust and UEQ hedonic appeal relative to the \textit{Answer-only} baseline, with no reliable separation between \textit{Full-trace} and \textit{Summary-trace}.

Across hypotheses, trace exposure reliably shifted the surface of the interaction -- trust and hedonic appeal -- but did not shift effort, usability, performance, or calibration in the predicted directions. The unifying account is that users engaged with the trace superficially, as a fluent artifact rather than as an object of evaluation. Where the preregistration anticipated comparison between the user's and the model's reasoning, participants instead treated the trace as a coherent rendering whose internal plausibility substituted for their own monitoring.

Across the full sample, every CAIMS subscale positively predicted overestimation and calibration error, replicating an inversion reported by \citet{FERNANDES2026108779}. Self-report instruments of (metacognitive-) AI literacy appear to track something other than the behavioral competencies they are meant to index: participants who report planning, monitoring, and evaluating their reasoning with AI are not the participants whose confidence tracks correctness. Our study adds to \citet{FERNANDES2026108779} that self-report measures of AI literacy seem increasingly unsuited to certifying metacognitive support in HAI interfaces.

\subsection{Complementarity, Overreliance, and Metacognitive support}
Reasoning traces are often marketed as transparency mechanisms assumed to help users understand why the model produced an answer and improve their judgment \cite{openai2024learning,openai2025monitorability,anthropic2025extended,anthropic2025think}. Our findings complicate this assumption. Full traces made the model's answer appear more inspectable and internally coherent, yet this additional information did not translate into better cognitive performance or calibration in the LSAT questions. Instead, it may have created an illusion of understanding \cite{fisher2021harder} due to the model's verbosity and apparent fluency, leading participants to trust the model's answer, consistent with the trust pattern predicted under H3.
These findings suggest that providing users with more information about how an LLM arrived at its answer does not necessarily support better reasoning, and may, in some cases, undermine it. Taken together, they reposition reasoning traces along three connected positions in HAI: complementarity, overreliance, and metacognitive support.

The \textit{complementarity} premise holds that humans and machines contribute distinct competencies whose joint output exceeds either alone -- i.e., \textit{synergy} \cite{vaccaro2024combinations}. However, recent evidence already qualifies this premise away from synergy toward \textit{augmentation}, where human-AI pairing merely exceeds human performance working alone \cite{vaccaro2024combinations,FERNANDES2026108779}. Our results extend the qualification to trace-mediated collaboration: contrary to H4, \textit{Full-trace} performance sat at the model-alone benchmark, indicating that trace exposure shifted the joint outcome toward the model's level rather than beyond it.  Qualitative patterns explain why. Only a minority of participants used the trace for calibrated comparison against their own reasoning, the discourse which complementarity presupposes \cite{FERNANDES2026108779}.%

This reallocation also reframes \textit{overreliance}. The cost-of-engagement account predicts that long, effortful explanations should reduce inappropriate reliance \cite{vasconcelos2023explanations}, yet the \textit{Full-trace} condition, which operationalizes that high-cost variant, produced the largest overestimation bias and the highest hedonic ratings, against the calibration improvement preregistered in H1 and in line with the trust elevation preregistered in H3. Our mediation analysis isolates a misattribution account: both trace conditions raised trust and hedonic appeal, but only hedonic appeal carried the indirect path to overestimation. This is inconsistent with the cost-of-engagement prediction \cite{vasconcelos2023explanations} that effortful traces reduce overreliance, and consistent with traces operating placebically when coherence outruns diagnosticity \cite{eiband2019impact}. 

The same mechanism explains why traces fail to support \textit{metacognition}. Articulate model reasoning has been treated as a candidate scaffold for users' planning, monitoring, and evaluation \cite{FERNANDES2026108779}, but contrary to H1, calibration bias was largest under full traces, and sensitivity indices did not differ across conditions: additional reasoning output did not increase the diagnosticity of users' confidence judgments and instead inflated overall self-evaluation. The corresponding null effects on task load (against H2) and on usability ratings (against H5) further indicate that the costs of full traces are not registered by users at the level of effort or usability and instead surface only at the level of calibration. Consistent with this, self-reported metacognitive engagement on the CAIMS positively predicted overestimation and calibration error rather than reducing them, indicating that users who believe they are monitoring their own reasoning are not the users whose confidence tracks correctness (also replicating \cite{FERNANDES2026108779}). A fluent and internally coherent rendering of reasoning renders user-side monitoring redundant, with model coherence substituting for the user's evaluation in line with \citet{alter2009uniting}. Thus, appropriate calibration, complementarity, and reliance in HAI with reasoning traces may depend on whether the interface protects the user's own reasoning from the model's fluency.

\subsection{Reasoning Traces as an Interface Artifact}
One key concern cuts across the data. Given that reasoning traces cannot serve as cues for correctness and, thus, support appropriate metacognitive calibration, their design value should lie at least in their capacity to serve model interpretability. Recent work, however, shows that chain-of-thought outputs diverge from the internal computations that produce a model's answer, with model reasoning and observed behavior coming apart \cite{10.5555/3666122.3669397,lindsey2025biology, barez2025chain}. 

Taking this view, users in the \textit{Full-trace} and \textit{Summary-trace} conditions may not be calibrating their decisions against interpretability cues derived from internal model states, but against a \textit{rendering} of reasoning produced for the user. The trace is thus best understood as an interface artifact whose user-centered persuasive properties are partly decoupled from its diagnostic ones. Indeed, recent work shows that AI explanations that provide \textit{faithful} interpretability information about a prediction model \textit{can} improve metacognition, particularly when they highlight divergences between human and AI logic \cite{von2025knowing}. This faithfulness problem, therefore, compounds the metacognitive one: users form increasingly confident judgments on the basis of cues that are themselves unreliable representations of model behavior. This echoes findings from cognitive psychology on the distinction between proxy and faithful explanations \cite{rozenblit2002misunderstood} and longstanding findings that fluent external information inflates perceived understanding without improving judgment \cite{alter2009uniting}. From this lens, reasoning traces in LLMs inherit both properties at once, and interpretations of such interfaces as transparency tools should be understood accordingly.%

Our qualitative data points to a second fundamental issue of reasoning traces in LLM interaction. Across both \textit{Full-trace} and \textit{Summary-trace}, participants used traces in a multitude of ways: to offload their own reasoning, to structure their own reasoning, as an anchor for calibration, or viewed them as a burden. Of these, only calibration would be genuine problem-solving with the potential for synergy in HAI. All other strategies yield following rather than engaging in discourse with the AI and thus bring parity with AI in terms of cognitive performance, as noted by \citet{vaccaro2024combinations}, resulting in AI augmentation, i.e., humans performing on par with AI.

\subsection{Design Implications}
Our findings suggest three directions for exposing reasoning traces to users. 

First, summaries preserved performance with marginally lower overestimation than full traces, while matching them in terms of trust and UX. Combined with qualitative evidence that verbose traces are skimmed, this suggests that verbose traces add little user benefit. Our qualitative findings highlighted the role of trace fluency in guiding human reasoning. Thus, at minimum, systems should aim to reduce the perceived fluency of traces and similar explanations, for example, by removing rhetorical features and anthropomimetic cues that imply human-like reasoning.

Second, given the concerns of faithfulness in traces and the decoupling of coherence and performance gains, design should be careful not to imply that visible and plausible traces certify the correctness of the answer. Thus, framing reasoning traces as windows into the inner workings of a model seems neither correct \cite{10.5555/3666122.3669397,lindsey2025biology, barez2025chain} nor useful for LLM users and must be toned down.%

Third, the qualitative data suggest that the real value of a trace may not lie in the trace itself but in the opportunity it creates for users to articulate their own reasoning before comparing it to the model's. From this perspective, what matters is less the format or even the quality or faithfulness of the trace but rather whether the interface elicits the user's own thinking in the first place. Simple scaffolds- 
prompting users to describe their approach, intuition, or tentative solution while the model is generating its response ("What do you think?"), may hold more promise as metacognitive interventions than refinements to the trace or the interface to traces themselves. For example, AI explanations can improve human metacognitive accuracy and AI-assisted decision-making when they reveal \textit{divergences} between human and AI reasoning (i.e., contrastive explanations) \cite{von2025knowing, buccinca2025contrastive, si2024large}. Thus, one can speculate that making the user form human reasoning traces in the first place may be as, or even more, important than designing reasoning traces from an HAI perspective \cite{FERNANDES2026108779}.

\subsection{Limitations and Future Work}

Our study has several limitations that bound the interpretation of its findings.

First, we rely on LSAT logical reasoning, which may not capture the diversity of real-world reasoning and likely overlaps with the model's training data. However, the models were far from ceiling on the retained items ($M = 5.00$ out of 10), and the mechanism we identified, fluent rendering inflating perceived understanding without improving judgment, is generic to settings where users must form a judgment without external verification. The same logic bounds questions of population and interface: we recruited US-based, English-fluent Prolific participants on a desktop layout, but the cognitive mechanism we investigate is established across populations and modalities. A follow-up would manipulate verifiability within a fixed task, contrasting items where ground truth is one click away (calculator, unit test, search) with items where it is not, to test whether traces support calibration only when an external verifier is present.

Second, we tested several models of one family with one rendering of full and summary traces, leaving the verbosity dose-response unmeasured. A methodological requirement compounds this: \textit{Full-trace} was instantiated with gpt-oss-20b, while \textit{Answer-only} and \textit{Summary-trace} used GPT-5. Items were pre-screened so that both models achieved the same mean performance at about 50\% on retained items, mitigating differences in model ability, but response style, error patterns, and surface fluency may still differ. The cleanest comparison in our design of \textit{Summary-trace} versus \textit{Answer-only}, both instantiated with GPT-5 is unaffected by this confound and supports our core claim that summaries preserve performance at the no-trace baseline while elevating trust and hedonic appeal. The \textit{Full-trace} effect should therefore be read as conditioned on an open-weight reasoning model that exposes verbose intermediate steps, which is also the user-facing reality of contemporary reasoning models. The design space has already converged on roughly the two patterns we contrast, and the underlying psychological mechanism is robust. Nevertheless, a more controlled follow-up would hold the producing model fixed and contrast full traces against programmatic summaries of those same traces, ideally also varying uncertainty marking against confident phrasing as in \citet{kim2024m}, to disentangle verbosity, fluency, and model identity.

Third, in \textit{Full-trace}, the model's reasoning was shown before participants could reveal its final answer via a button press, whereas in \textit{Summary-trace}, the trace and answer appeared together. Trace condition is therefore confounded with the engagement requirement. This design choice was made (and also preregistered) to ensure participants encountered the trace at all, and if anything, should have favored deliberation: forced engagement gave participants an opportunity to form their own answer before being anchored by the model's. The \textit{Full-trace} performance impairment therefore arises despite a feature designed to support critical engagement, making the effect shown rather conservative. A follow-up crossing trace verbosity with engagement requirement (forced reveal versus passive presentation) may decompose trace content from pre-answer deliberation. Nevertheless, given the low engagement in the qualitative data, it seems unlikely that this has led to shallow engagement with the long traces. 

Fourth, the ten-item battery limits the metacognitive measures that can be reliably estimated; signal-detection-based indices such as meta-d' and M-ratio \cite{rahnev2025comprehensive} require substantially more trials per participant than our design provides, even under hierarchical Bayesian estimation. We addressed this by reporting four convergent sensitivity indices, confidence gap, AUROC2, calibration error, and Brier score, and the consistency of the sensitivity null across all four measures would be unlikely if it were driven by noise in any single index. Per-participant noise also biases against detecting between-condition differences, making our positive findings conservative. A within-subjects extension with a much larger battery would enable HMeta-d estimation and link metacognitive efficiency to the interaction strategies identified in our qualitative typology. However, for LLM interaction studies, running studies with 100-400 trials to estimate signal-detection parameters of confidence is uncommon. 

Fifth, like \citet{FERNANDES2026108779}, we observe that AI assistance can boost performance without commensurate gains in metacognitive accuracy; reasoning traces appear to deepen, rather than repair, this effect. Because we measure trust, confidence, and bias jointly, and our trust and hedonic mediators correlate strongly, we cannot fully separate calibration from these effects, though our partition of variance shows that hedonic appeal carries the unique indirect path to overestimation. A more controlled follow-up would impose a disagreement schedule in dialog, forcing the model to push back on the user a fixed fraction of the time, independent of correctness, and to ask whether users update symmetrically when right or wrong. A complementary manipulation of fluency at fixed length would isolate the hedonic component from trust directly.

Sixth and finally, our design speaks to in-the-moment calibration, not to long-term learning or deskilling~\cite{bastani2024generative}. A longitudinal study could examine prolonged trace exposure alongside unaided metacognition to test how sustained exposure to confident traces erodes users' ability to know when they are wrong on tasks they have always been able to do.

\section{Conclusion}

In our preregistered between-subjects study on LSAT-style reasoning ($N=559$), more reasoning output did not support interaction: full traces impaired performance relative to the answer-only baseline and produced the largest overestimation gap, while summaries merely preserved baseline performance. Both trace conditions, however, raised trust and user experience. Reasoning traces are therefore best understood as user-facing interface artifacts rather than transparent windows into model cognition, and metacognitive calibration is unlikely to emerge from the trace itself. Calibrated reliance in HAI requires scaffolding users' own reasoning before exposing them to the models', shifting the design question from how to render the trace to how to support the user's thinking around it.



\bibliographystyle{ACM-Reference-Format}
\bibliography{references}

\appendix

\section{Nasa-TLX Questionnaire}

\begin{table}[H]
\centering
\caption{Subjective workload across conditions, measured by the NASA Task Load Index (NASA-TLX)}
\begin{tabular}{lccc}
\hline
\multicolumn{1}{c}{Condition} & \textit{Answer-only} M(SD)    & \textit{Full-trace} M(SD)      & \textit{Summary-trace}   M(SD)    \\ \hline
Nasa TLX                      & 43.85 (13.09) & 45.19 (14.39) & 45.58 (13.51) \\
Mean Mental Demand            & 71.42 (22.72) & 73.66 (22.89) & 74.71 (21.15) \\
Mean Physical Demand          & 26.95 (25.89) & 33.58 (28.22) & 39.37 (31.62) \\
Mean Temporal Demand          & 26.47 (19.82) & 31.37 (22.09) & 27.83 (19.40) \\
Mean Performance              & 32.75 (24.62) & 33.44 (27.80) & 31.35 (27.03) \\
Mean Effort                   & 67.35 (23.12) & 69.40 (24.88) & 71.46 (21.99)  \\
Mean Frustration              & 38.16 (29.62) & 29.70 (25.96) & 28.78 (25.03)   \\
\hline
\end{tabular}
\label{tab:nasa}

\raggedright \small \textit{Note.} Means and standard deviations are reported on a 0-100 scale. The overall NASA-TLX score was computed as the mean of the six subscales. Overall workload did not differ significantly across conditions.
\end{table}

\section{Qualitative Analysis}

\subsection{Full-trace Condition}
\begin{table}[H]
\centering
\caption{Theme groups based on the type of interaction with AI \textit{Full-trace} condition.}
\begin{tabular}{cll}
\hline
Group & \multicolumn{1}{c}{Label}                                            & Number of participants (\%)        \\ \cline{1-3}
1     & Used AI to obtain answer and cognitive offloading  & 65 (35.3\%)     \\
2     & AI as guided reasoning aid and decision support    & 57 (31.1\%)   \\
3     & Independent verification and calibrated comparison & 38 (20.8\%)   \\
4     & Reasoning trace burden                           & 19 (10.4\%)     \\
5     & Unclear or inconclusive                           & 3 (1.6\%)     \\
6     & AI not useful                                    & 1 (0.5\%)    \\
\hline
\end{tabular}

\label{tab:qualitative_C2}

\raggedright \small \textit{Note.} Groups are mutually exclusive. Percentages are computed relative to the condition sample size.
\end{table}

\subsection{Summary-trace Condition}

\begin{table}[H]
\centering
\caption{Theme groups based on the type of interaction with AI \textit{Summary-trace}}
\begin{tabular}{cll}
\hline
Group & \multicolumn{1}{c}{Label}                            & \multicolumn{1}{c}{Number of participants (\%)} \\ \hline
1     & AI as summary-based reasoning aid and decision support & 74 (39.2\%)                                     \\
2     & Independent verification and calibrated comparison     & 54 (28.6\%)                                     \\
3     & AI as primary answer source and low interaction use    & 52 (27.5\%)                                     \\
4     & Unclear or inconclusive use              & 9 (4.7\%)  \\           
\hline
\end{tabular}

\label{tab:qualitative_C3}

\raggedright \small \textit{Note.} Groups are mutually exclusive. Percentages are computed relative to the condition sample size.
\end{table}

\end{document}